\documentclass[pre,floats,superscriptaddress,usenames]{revtex4}

\usepackage{amsmath}
\usepackage{graphicx}
\newcommand{\be}{\begin{equation}} 
\newcommand{\ee}{\end{equation}}
\newcommand{\bea}{\begin{eqnarray}}   
\newcommand{\eea}{\end{eqnarray}}

\newcommand{\rr}{{\bf r}}
\newcommand{\rrt}{{({\bf r},t)}}
\newcommand{\NN}{{\bf \nabla}}
\newcommand{\FF}{{\bf F}}

\newcommand{\vv}{{\bf v}}

\newcommand{\fa}{f^{\alpha}}
\newcommand{\na}{n^{\alpha}}

\newcommand{\uu}{{\bf u}}
\newcommand{\uua}{{\bf u}^{\alpha}}

\newcommand{\uai}{u^{\alpha} _i}
\newcommand{\uaj}{u^{\alpha }_j}

\newcommand{\ma} {m^{\alpha}}

\begin{document}

\date{\today}

\title{Electro-osmotic flow in  coated nanocapillaries: a theoretical investigation}

\author{Umberto Marini Bettolo Marconi}

\address{Scuola di Scienze e Tecnologie, 
Universit\`a di Camerino, Via Madonna delle Carceri, 62032, Camerino, INFN Perugia, Italy}

\author{Michele Monteferrante}

\address{CNR-ICRM, Consiglio Nazionale delle Ricerche, 
Istituto di Chimica del Riconoscimento Molecolare, Universit\`a La Sapienza, 
Via Mario Bianco, 20131 Milan, Italy.}

\author{Simone Melchionna}

\address{Istituto Processi Chimico-Fisici, Consiglio Nazionale delle Ricerche, Italy}

\vspace{0.6cm}

\begin{abstract}

Motivated by recent experiments, 
we present a theoretical investigation of how the electro-osmotic flow occurring in a  capillary 
is modified when its charged surfaces are coated by charged polymers.
The theoretical treatment is based  on a three dimensional model consisting of a ternary fluid-mixture, representing the solvent and  two species for the ions,
confined between two parallel charged plates decorated by a fixed array of scatterers representing the polymer coating.
The electro-osmotic flow, generated by a constant electric field applied in a direction parallel to the plates,
is studied  numerically by means of  Lattice Boltzmann simulations.
In order to gain further understanding we performed a simple theoretical analysis by
extending the Stokes-Smoluchowski equation to take into account the porosity induced by the polymers
in the region adjacent the walls.
We discuss the nature of the velocity profiles  by focusing on the competing effects
of the polymer charges and the frictional forces they exert. We show evidence of the flow 
reduction and of the flow inversion phenomenon
when the polymer charge is opposite to the surface charge.
By using the density of polymers and the surface charge as control variables, 
we propose a phase diagram that discriminates the direct and the reversed flow regimes
and determine its dependence on the ionic concentration.

\end{abstract}

\maketitle

\section{Introduction }

Electrokinetic phenomena of fluids under conditions of extreme confinement are important to micro and nanofluidics 
and have a variety of applications ranging from fabrication of efficient nanotechnological devices, electrochemical energy storage,
electrokinetic energy conversion, up to 
  biomedical applications, such as 
separation and analysis of biological molecules and
 molecule delivery and sensing \cite{bruus2008theoretical,berthier2010microfluidics,kirby2010micro,nguyen2002fundamentals,masliyah2006electrokinetic,
tabeling2014physics}.
 What makes micro-sized channels attractive is the large surface to volume ratio,
so that the surface has a greater impact and some new phenomena arise, opening the possibility 
to develop new fluidic functionalities, since 
decreasing the scales increases the sensitivity of analytic techniques which are used in Lab on a chip (LOC) devices
\cite{rotenberg2013electrokinetics,kontturi2008ionic}.

The present paper investigates the effect of coating the inner walls by polymers on the electro-osmotic flows (EOF).
In standard  electroosmosis  the motion of ions and their surrounding water
molecules,   induced by an  applied electric field
parallel to the charged walls of a capillary, generates a flow that extends outside
the Debye electric double layer (EDL).
The presence of a non uniform velocity profile  associated with the EOF may represent a 
problem in capillary electrophoresis or in microfluidic devices used in protein analysis,  because it increases dispersion and reduces resolution. On the contrary, in capillary electrochromatography the flow must be enhanced to produce
high throughputs.
It has been demonstrated by experiments, computer simulations \cite{hickey2011influence,hickey2012computer} and theoretical arguments \cite{monteferrante2014electroosmotic}   that a modification of the chemical composition of the EDL
by  coating the walls with polymers leads  to a consistent reduction of the mass flow and even to a reversal of the electroosmotic
current \cite{hickey2011influence}. 
This modification of the flow may occur for two reasons: the first is the drag force exerted by the polymer beads on the electrolyte
solution and the second is the electric field originated by the polymer charges. 
Danger and coworkers \cite{danger2007control}
 observed that decreasing the EOF optimizes the resolution and analysis
time in the separation of peptide mixtures in coated capillaries and investigated which polymers
were more efficient to realize such a situation. 

In spite of the large amount of experimental work \cite{doherty2002critical,znaleziona2008dynamic,horvath2001polymer,chiari2000new} on the mechanisms  controlling the EOF in polymer coated capillaries,
a complete understanding of the problem is still missing \cite{shendruk2012electrophoresis}. 
The scaling theory of Harden et al. \cite{harden2001influence} focused  on the interplay between 
the deformation of the adsorbed polyelectrolytes and the fluid motion. A series of 
Molecular dynamics (MD) simulations gave support to their predictions regarding the dependence of the polymer length
on the coatings.  
The influence of the polymeric structure 
and of the solvent  on the EOF was also investigated
at length scales smaller than the Debye length  \cite{qiao2007modulation,cao2010electroosmotic,cao2012modulation} . 
Slater and coworkers \cite{tessier2006modulation,hickey2009molecular} 
found by MD  that the direction of the EOF can reverse in the region near the walls. 

We propose here an alternative simulation approach to determine the EOF in coated capillaries,
based on the Lattice Boltzmann method (LBM) \cite{benzisuccivergassola}
and is supported by  a straightforward theoretical treatment of the current modulation.
We believe that our numerical approach can be generally employed for this type of systems
and is complementary to particle-based simulations, the latter being 
computationally more expensive such that only a limited range of
length scales, electrolyte concentrations and polymer coatings can be explored. 
In fact, MD provides information at the molecular level at the price of
the heavy computational effort required to track individual molecules and, 
in order to reach statistical accuracy for electrolytic solutions where the ionic densities 
are several orders of magnitude smaller than solvent density, demands for 
expensive ensemble averaging procedures \cite{karniadakis2006microflows}.

 Our modeling of the polymer coating is idealized and neglects the 
 fact that polymers with one end fixed to the surface  tend to  deform due to the motion of the fluid.
The polymer layer is assimilated  to a region populated by an assembly of fixed random  charged obstacles exerting
 a drag force on  the fluid.  A standard description of such a situation is represented
by the  phenomenological  Brinkman equation \cite{looker2006electrokinetics},
often used to model the behavior of the fluid velocity, $\uu$, through highly permeable porous media
as:
\be
\eta_{eff} \nabla^2 \uu=\nabla p +\frac{\eta}{k} \uu
\label{darcy}
\ee
where $\nabla p$ is the pressure gradient, $\eta$ and $\eta_{eff}$ are the standard and  effective dynamic viscosity, respectively,   $k$ is the permeabilty of the medium, with
  $\eta_{eff}=0$ corresponding to the Darcy equation.

In the following we will determine the properties of the EOF in slit-like capillaries with coating 
represented at finer level with respect to the description of eq. \ref{darcy}, 
by using the LBM approach:
a formula equivalent to eq. \ref{darcy} will be obtained    
from the  one-particle
phase space distributions of the individual species as basic variables through an averaging procedure. The fluid is described as a
a ternary mixture comprising a solvent species and two ionic species of opposite charge, as 
recently done in ref. \cite{melchionna2011electro}.  The LBM provides a
 statistical description of the system under study, allows to compute density profiles and flow fields and embodies correctly the hydrodynamics and the electrostatics. 
To this purpose, we consider
the simplest version of  the LBM for electrolytes which is an extension of the  Bhatnagar-Gross-Krook    (BGK) \cite{bhatnagar1954model} relaxation-time model.
It neglects the short range structure of the fluid and the non ideal gas features of the fluid besides
the long range electrostatic interactions, but also displays undeniable advantages such as the simplicity of programming,
speed of calculation and allowing to explore arbitrary
channel geometries.

This paper is organised as  follows: 
in \ref{Model} we illustrate the
 Boltzmann-like  transport approach
describing a ternary mixture comprising a solvent species and two ionic species of opposite charge.
The non electrostatic part of the collision terms is treated within a single relaxation time
approximation of the BGK type \cite{bhatnagar1954model}.
We derive the balance equations for the momenta and the densities
 and recover the 
structure of equations for the charge, the mass and  the momentum 
predicted by the macroscopic models.
In \ref{Analytics}
in order to gain further insight we perform a reduction of the Boltzmann coupled equations to a Stokes-Smoluchowski
formulation  and obtain the electric potential profile within the pore in the Debye-Huckel approximation \cite{masliyah2006electrokinetic}
and show analytically how the mass current can be modulated 
by changing the properties of the polymer adsorbed on the surfaces.  
We study the resulting  velocity field as a function of the relevant parameters. 
In sections \ref{Numerics} and \ref{Numerics2} we validate such results by solving numerically the LBM equations 
and treat the electric potential using the Poisson equation. 
We compare the numerical and theoretical results for the velocity profile and the potential
and show that in the limit of small surface charge  densities the agreement regarding the global  properties of the system 
are satisfactory, whereas for larger densities the deviation is due to the non-linear nature of the self-consistent
electric potential.
Finally, in  section \ref{Conclusions} we make some conclusive remarks and considerations.

\section{Theoretical background}
\label{Model}

 Let us consider a ternary mixture and denote the components by the superscript $\alpha=0,-,+$, where $0$ identifies the solvent,
$-$  and $+$ indicate  the anions and the cations, respectively. 
The three types of particles have masses $\ma$, charges $z^\alpha e$ (with $z^0=0$), expressed in 
units of the electronic charge $e$, and are free to move in a slit-like 
channel whose parallel walls coincide with the planes at $z=\pm w$, are impenetrable and carry a fixed surface charge
of density per unit area $\Sigma $, which is negative in the cases here considered.

Our goal is to determine the relevant dynamical  observables of the system,
such as the partial number density $\na\rrt$ and local velocity $\uu^\alpha\rrt$ starting from the
one-particle phase space distributions $\fa(\rr,\vv,t)$
by performing the following projections in velocity space:
\be
\left( \begin{array}{ccc}
\na(\rr,t)  \\
 \na(\rr,t)\uua(\rr,t)  \end{array} \right) =
\int d\vv
\left( \begin{array}{ccc}
1   \\
 \vv    \end{array} \right)  f^\alpha(\rr,\vv,t).
\label{colonna}
\ee
From these quantities one derives the local charge distribution:
\be
\rho(\rr,t)=e\sum_\alpha z^\alpha\na(\rr,t)\, ,
\ee
the global number density: 
\be
n(\rr,t)=\sum_\alpha \na(\rr,t) 
\ee
and the average barycentric velocity of the fluid:
\be
\uu(\rr,t)=\frac{\sum_{\alpha}\ma\na(\rr,t)\uua(\rr,t)}
{\sum_{\alpha}\ma\na(\rr,t)} \, .
\ee
Hereafter, in the notation we will drop the dependence $(\rr,t)$ for these fields for the sake of clarity and
also assume the temperature to be constant.

 We now consider the  evolution equation  of the $\fa$'s (see  reference \cite{marini2012charge}  for details) :
\bea
\frac{\partial}{\partial t}\fa(\rr,\vv,t)
&+&
\vv\cdot\NN \fa(\rr,\vv,t)
+\frac{\partial}{\partial \vv}
(\frac{\FF^{\alpha}(\rr,\vv)}{\ma}\cdot
 \fa(\rr,\vv,t))
\nonumber\\
&&
=
-\omega[ \fa(\rr,\vv,t)- \psi^{\alpha}(\rr,\vv,t)]
\nonumber\\
&&
+\frac{e z^\alpha }{\ma}\NN \phi(\rr)\cdot
\frac{\partial}{\partial \vv} \fa(\rr,\vv,t)  \, . \nonumber\\
\label{evolution}
\eea
The left hand side of  eq. \ref{evolution} represents the streaming contribution to the evolution, while the right hand side 
contains a  BGK relaxation term representing the effect 
of the molecular collisions tending to restore on time scale $\omega^{-1}$ the local equilibrium distribution represented by:
\be
\psi^{\alpha}(\rr,\vv,t)=\na[\frac{\ma}{2\pi k_B T}]^{3/2}\exp
\Bigl(-\frac{\ma(\vv-\uu)^2}{2 k_B T} \Bigl)
\label{psia}
\ee
with $k_B$ is the Boltzmann constant. 

The last term in the r.h.s. of eq. \ref{evolution}  describes  the electrostatic coupling among the ions, 
treated within the Vlasov mean-field  approximation, where one assumes that
the electrostatic potential $\phi$ is the solution of the Poisson equation:
\be
\nabla^2 \phi= -\frac{\rho_{tot}}{\epsilon}
\label{poisson}
\ee
where $\rho_{tot}$ is the sum of the mobile ionic charge density $\rho$ 
and the density due to fixed  charges $\rho_f(\rr)$ distributed within the system.
The confining walls impose the vanishing of the densities $\na$ for $|z|>w$
and of the velocities $\uua$ at $z=\pm w$ according to the no-slip boundary condition.
The walls also contribute 
to the electric potential $\phi$, due to their surface charges.
The fixed surface charges appear through the
Neumann boundary conditions
on the gradient of the electrostatic potential:
\be
\hat n \cdot \nabla \phi|_{\rr \in S(\rr)}=-\frac{\Sigma(\rr)}{\epsilon}
\label{Poissonboundary}
 \ee
where $\hat n$ is the local normal to the surface, $S(\rr)$.
The associated value of the potential at the wall is: 
 \be
 \phi|_{\rr \in S(\rr)}\equiv\zeta .
\label{Dirichlet}
 \ee 
We model the interaction between the polymers and the fluid by taking into account that
the polymers are adsorbed irreversibly  on the surface and exert two kinds of forces 
on the fluid:
a) a drag force in the region where
the mobile particles hit the polymer and their momenta are reduced; 
b) an electric force, because the polymer charges interact with the ions of the solution
and modify the structure of the EDL. 
The polymer coating is idealized as 
 a region of thickness $\delta=(w-h)$ and volume $V_s$ adjacent to the wall and containing $N_s$ fixed charges
described by the density field
\be
\rho_{f}(\rr)=e z_s \sum_{k=1}^{N_s}   \delta(\rr_k-\rr),
\ee
where $\rr_k$ are the positions of the charges and $z_s$ their valences,
and act as sources of the potential $\phi(\rr)$ via eq. \ref{poisson}.
The drag force is assumed to be proportional to the velocity of the moving particles
 and to the density of obstacles:
\be
\frac{\FF_{drag}^\alpha(\rr,\vv)}{\ma}=- k_p \sum_{k=1}^{N_s} \vv\delta(\rr_k-\rr)
\label{dragforce}
\ee
where $k_p$ is a measure of its strength.
eq. \ref{evolution} reproduces the correct hydrodynamic behavior,
in fact, by integrating w.r.t. the velocity in eq. \ref{evolution} we obtain the conservation law for the
particle number of each species:
\be
\frac{\partial}{\partial t}\na +\nabla\cdot \Bigl(\na \uu\Bigl)+
\nabla\cdot \Bigl(\na(\uua- \uu)\Bigl)=0 ,
\label{continuity2}
\ee
where the last term in eq. \ref{continuity2}
is the so-called dissipative diffusion current,
measuring the drift of the $\alpha$-component with respect to the barycentric velocity.
After multiplying by $\vv$ and integrating eq. \ref{evolution}  w.r.t. $\vv$, we obtain
the balance equation for the density of momentum of the species $\alpha$:
\bea
&&
\ma\frac{\partial}{\partial t}[\na\uaj]+ 
\ma\nabla_i \Bigl(\na \uai \uaj
-  \na(u^{\alpha}_i-u_i)( u^{\alpha}_j-u_j)\Bigl)
\nonumber\\
&& 
=-\nabla_i  {\cal P}_{ij}^{\alpha} -k_p  \ma \sum_{k=1}^{N_s} \delta(\rr-\rr_k) u^\alpha_j \na
\nonumber\\
&&
 -e z^\alpha\na\nabla_j \phi-\omega \ma\na (u^\alpha_j-u_j) \nonumber\\
\label{momentcomponent}
\eea
where
\be
{\cal P}_{ij}^{\alpha}=\ma\int d\vv (v_i-u_i)(v_j-u_j)\fa(\rr,\vv,t)
\label{pressurekin}
\ee
represents the kinetic contribution of component $\alpha$ to the pressure tensor  \cite{marconi2011dynamics}.
In eq. \ref{momentcomponent} and in the following, the convention of summing over
repeated indices is assumed.


The balance of the total moment of the fluid follows immediately by summing
eq. \ref{momentcomponent} over the three species:

\bea
&&\rho_m \partial_{t}u_j+ \rho_m u_i\nabla_i u_j+\nabla_i P_{ij}- 
\nonumber \\
&&
\rho_m k_p  u_j \sum_{k=1}^{N_s} \delta(\rr-\rr_k) -
\sum_\pm e z^\pm n^\pm\nabla_j \phi=0  
\nonumber \\
\label{globalmomentumcont}
\eea
where $\rho_m=\sum_\alpha \ma \na$ is the local mass density and
$
{\cal P}_{ij}=
\sum_\alpha {\cal P}_{ij}^{\alpha}
$ is the total (kinetic) pressure.
By a standard analysis  \cite{marconi2011multicomponent} 
for species of equal masses ($m_\beta \equiv m$), one obtains the mutual diffusion coefficient 
 $D=\frac{k_B T}{m \omega}$,
the dynamic shear viscosity $\eta=\frac{k_B T}{m \omega} \rho_m$ and the
total kinetic pressure:
\be
{\cal P}_{ij}=k_B T n \delta_{ij}-\eta \Bigl( \frac{\partial u_i}{\partial x_j}+ \frac{\partial u_j}{\partial x_i}-\frac{2}{3}
 \frac{\partial u_k}{\partial x_k} \delta_{ij}  \Bigl)
 \label{pressure}
 \ee


\section{Analytic treatment}
\label{Analytics}
The presence of charges on the walls and on the fixed polymers gives rise to the formation of an EDL
along the $z$ direction. Under the action of an electric field parallel to the walls induced by the presence of  two electrodes, 
the excess positive charge in the EDL generates a flow of positive ions
towards the cathode and their motion is transmitted by shear forces to the rest  of the fluid in the capillary.

Before embarking on the full numerical approach to the problem, it is useful to 
study the solution of the electro-hydrodynamic equations  eq. \ref{poisson} and eq. \ref{globalmomentumcont} by
 using the linearized Debye-Huckel theory to treat the EDL 
and the  Stokes approximation to describe  the creeping flow regime, commonly realized in micro 
and nanofluidics set-ups. In this limit,
 the non-linear terms  in the velocity in eq. \ref{momentcomponent} can be neglected being much smaller than the viscous term and will be dropped in the rest of the paper.
To proceed analytically
we average over the distribution of charged obstacles, assuming that they are uniform within the two layers
each of volume  $V_s$  where they are located:
$\frac{N_s}{V_s}= \langle  \sum_{k=1}^{N_s} \delta(\rr_k-\rr)\rangle$.
 Such an averaging leads to smooth densities and velocity profiles compared to the noisy profiles
observed in the simulations, but allows to 
exploit the statistical invariance with respect to translations in directions parallel to the 
walls.

The resulting stationary state of
the total  momentum eq. \ref{globalmomentumcont}, averaged over the polymer distribution, 
is a solution of the following equation for the fluid velocity:
\be
\eta  \frac{\partial^2 u_x}{\partial z^2}=\nabla_x p
-E_x\sum_\pm e z^\pm n^\pm+\gamma\rho_m \Theta(z) u_x =0
\label{globalmomentumcontc2}
\ee
where we introduced the friction constant $\gamma=k_p \frac{N_s}{V_s}$ and  
the driving electric field $E_x=-\nabla_x \psi(x)$  parallel to the walls.
The function
$\Theta(z)=(\theta(w-h-z)+\theta(z-w+h))$,  constructed as the sum of two Heaviside distributions,
describes the presence of the polymers in two slabs adjacent the walls
and  $\nabla_x p$ indicates
the  term   $k_B T\nabla_x n$   associated with a possible pressure gradient along the $x$ direction.
We also  separate the pressure tensor into its dissipative and non dissipative contributions 
by using  eq. \ref{pressure}.
Under steady conditions the  density distributions are constant along planes parallel to the walls and, neglecting
steric effects due to the short range repulsive forces between molecules, can be determined
with the help of eq. \ref{poisson} and by  assuming the local equilibrium condition.
One obtains the following self-consistent Poisson-Boltzmann equation
for the electrostatic  potential $\phi$   between the plates:
 \be
\frac{d^2 \phi}{d z^2}=\frac{e n_b}{\epsilon} \sinh{\frac{(e  \phi)}{k_B T}}-\frac{e z_s n_s}{\epsilon}\Theta
\approx k_D^2 \phi- \frac{e z_s n_s}{\epsilon}\Theta,
\label{helmholtz}
\ee
where $n_b$ is the density of the single ionic species in the bulk and
$k_D$ is the inverse Debye length:
 \be
 \lambda_D=k^{-1}_D=\sqrt{\frac{\epsilon k_B T}{2 e^2 n_b}}.
 \ee
Notice that
in the r.h.s. of eq. \ref{helmholtz} we separated the source into two contributions,
the first taking into account the mobile charges and the second the charge of the obstacles. 
In the analytic treatment we 
employ the linear Debye-Huckel approximation for the electrostatic potential
corresponding to the second approximated equality in eq. \ref{helmholtz}, but in the numerical LBM treatment
we consider  the fully non-linear Poisson-Boltzmann equation. 
 The solution of eq. \ref{helmholtz}  with the properties of being continuous together with its first derivative
 at $z=\pm h$ and corresponding to the surface charge density $\Sigma$  at $z=\pm w$, reads:
 \begin{equation}
\phi(z) = \begin{cases} 
A \cosh(k_D z) +\frac{e z_s n_s}{\epsilon k_D^2}(1-\sinh(k_Dh)e^{-k_D|z|}) & \mbox{if }   w \ge |z| \ge h \\ 
 \tilde A \cosh(k_D z)      &  \mbox{if }   h \ge |z| \ge 0  
 \end{cases}
 \label{explicitrep}
\end{equation}
 where 
 \bea
  A&=&\frac{1}{ \sinh(k_D w)} \Bigl(\frac{\Sigma}{\epsilon k_D} -\frac{e z_s n_s}{\epsilon k_D^2} \sinh(k_Dh)e^{-k_Dw}\Bigl)  
 \\
  \tilde A&=& A+\frac{e z_s n_s}{\epsilon k_D^2} e^{-k_D h} \, .
\label{plicitrep}  
\eea   
    In order to obtain the modified  Smoluchowski equation for the coated capillary one must take into account the fact
that  also the polymer charge contributes to the Laplacian of the electric potential. 
Thus, 
we rewrite eq. \ref{globalmomentumcontc2} as
   \be
 \eta \frac{\partial ^2}{\partial z^2}u_x= \nabla_x p +\gamma\rho_m \Theta u_x
 + \epsilon E_x \frac{\partial ^2}{\partial z^2} \phi +  e  z_s E_x n_s \Theta 
 \, .
 \label{stokes1}
  \ee 
 The first two terms in the r.h.s. of  eq. \ref{stokes1} correspond to those featuring in the  r.h.s. of  
 Brinkman equation of porous media  eq. \ref{darcy}.   
  After substituting the explicit representation eq. \ref{explicitrep} of $\phi$ we recast the Smoluchowski equation as:
   \bea
 \left\{
 \begin{array}{l} 
 \frac{\partial ^2}{\partial z^2}u_x(z) 
 =\frac{\epsilon E_x k_D^2}{\eta} \tilde A \cosh(k_D z)  +\frac{\nabla_x p}{\eta}     \,\,\,\,\,\, \mbox{if }   h \ge |z|   \\
 \frac{\partial ^2}{\partial z^2}u_x(z) -\beta^2 u_x(z)
 = \frac{\epsilon E_x}{\eta} k_D^2 A \cosh(k_D z)   \\
\;\;\;\;\;\;\;\;\;\;\;\;\;\;\;\;\;\;\;\;\;\;\;\;\;\;\;\;\;\;\;\;\;\;\;\;\;\;  
- \frac{e z_s E_x n_s }{\eta}\sinh(k_D h)e^{-k_D |z|} \\
\;\;\;\;\;\;\;\;\;\;\;\;\;\;\;\;\;\;\;\;\;\;\;\;\;\;\;\;\;\;\;\;\;\;\;\;\;\;  
+ \frac{e z_s E_x n_s +\nabla_x p}{\eta} 
\;\;\;\;\;\;\;\;\;\;\;\;  
\,\,  \mbox{if }   w \ge |z| \ge h  \nonumber \\  
  \end{array}
\right .  
   \eea  
 where we defined the new characteristic inverse length, $\beta$, as
 \be
 \beta=\sqrt{\frac{\gamma}{\eta} \rho_m}  \,\,,
 \ee 
 related to the permeability, $k$, by $k=\beta^{-2}$.

   The solution with the properties that the fluid velocity  and its first derivative are continuous
 at $z=\pm h$ and  vanishes at the walls, $z=\pm w$, can be written in the following form:
 \bea
 \left\{
 \begin{array}{l}
 u_x(z)=G \cosh(k_D z)+F+\frac{\nabla_x p}{\eta} \frac{z^2}{2}   \,\,\,\,\,\, \mbox{if }   h \ge |z| \ge 0   \\ 
u_x(z)= C \cosh(\beta z)+D \cosh(k_D z) +H e^{-k_D |z|}+B e^{-\beta |z|} 
\\
\;\;\;\;\;\;\;\;\;\;\;\;\;  
-  \frac{e z_s E_x n_s+\nabla_x p }{\eta \beta^2} 
\;\;\;\;\;\;\;\;\;\;\;\;\;\;\;\;\;\;\;\;\;\;\;  
\,\,\,\,\,\,\,
\mbox{if }   w \ge |z| \ge h 
\end{array}
\right .
\label{vsolution}
\eea
  and the constants $D,C,G,F,H$ are determined by the set of equations:
 \bea
 \left\{
 \begin{array}{l}
 D=\frac{\epsilon E_x}{\eta}\,  \frac{k_D^2}{k_D^2-\beta^2}  A \, ,
 \\
 H=-\frac{ E_x}{\eta}\,  \frac{ ez_s n_s}{k_D^2-\beta^2}  \sinh(k_D h) \, ,
  \\
  G=\frac{\epsilon E_x }{\eta } \tilde A
\\  
 C \cosh(\beta w)+ B e^{-\beta w}=-D \cosh(k_D w) -H e^{-k_D w}
 \\
\;\;\;\;\;\;\;\;\;\;\;\;\;\;\;\;\;\;\;\;\;\;\;  
\;\;\;\;\;\;\;\;\;\;\;\;\;\;\;\;\;\;\;\;\;  
     + \frac{e z_s E_x n_s+\nabla_x p }{\eta \beta^2}
  \\
   C
  \beta \sinh(\beta h) -B \beta e^{-\beta h} =-k_D D \sinh(k_D h) +k_D e^{-k_D h} H
  \\
\;\;\;\;\;\;\;\;\;\;\;\;\;\;\;\;\;\;\;\;\;\;\;  
\;\;\;\;\;\;\;\;\;\;\;\;\;\;\;\;\;\;\;\;\;\;\;\;\;  
  + k_D G \sinh(k_D h)+\frac{\nabla p}{\eta}h
 \\
   F=C  \cosh(\beta h)+ D \cosh(k_D h) +H e^{-k_D h}+B e^{-\beta h}
  \\
\;\;\;\;\;  
  -G \cosh(k_D h)- \frac{e z_s E_x n_s+\nabla_x p }{\eta \beta^2}-\frac{1}{2}\frac{\nabla p}{\eta} h^2 \,.
  \end{array}
\right .  
\nonumber\\
  \eea  
  The volumetric flux, neglecting the variations of the densities, is given by 
$I_M=\rho_m \int_{-w}^w dz u_z(z)= \rho_m \Phi$
where
  \bea
  \Phi &=& 
  \frac{2 G}{k_D} \sinh(k_D h)+2 F h+2 \frac{C}{\beta} [\sinh(\beta w)-\sinh(\beta h)]
  \nonumber\\
  &+&  
  2 \frac{D}{k_D} [\sinh(k_D w)-\sinh(k_D h)]   
  \nonumber\\
  &-&  
 2 \frac{B}{\beta} [e^{-\beta w}-  e^{-\beta h}]   
-2 \frac{H}{k_D} [e^{-k_D w}-  e^{-k_D h}]   
  \nonumber\\
  &-&  
 2 \frac{e z_s E_x n_s+\nabla_x p }{\eta \beta^2}(w-h)+\frac{ \nabla p}{3 \eta}h^3 \,. \nonumber\\
     \eea
 Some simple limits can be analyzed:
 \begin{itemize}
 \item
 when the polymer charge vanishes and $k_D h\gg 1$ (thin EDL regime)  the velocity at midpoint is 
  \bea
 u_x(0)
&\approx&
 -\frac{E_x\Sigma}{\eta k_D} \frac{\cosh(\beta h)}{\cosh(\beta w)}
 \nonumber\\
 && -\frac{\nabla p}{\eta}\Bigl(\frac{h^2}{2}+\frac{1}{\beta^2}(1-\frac{\cosh(\beta h)}{\cosh(\beta w)})\Bigl)
 \nonumber\\
 & \approx &
  \begin{cases} 
 -\frac{E_x\Sigma}{\eta k_D}  -\frac{\nabla p}{\eta}\frac{w^2}{2}  & \mbox{if }   \beta h \ll 1 \\ 
  -\frac{E_x\Sigma}{\eta k_D} e^{-\beta(w-h)} -\frac{\nabla p}{\eta} \frac{h^2}{2}  &  \mbox{if }   \beta h\gg 1
 \end{cases}
 \nonumber\\
 \label{vmidopint1}
\eea
In the case $\beta h\ll 1$ one recovers the standard EOF plus the pressure induced flow in a slit
having the full width $2w$; in the case $\beta h\gg 1$ the EOF is exponentially suppressed at midpoint, while
the pressure induced flow corresponds to a slit having reduced width $2 h$.
\item
If the polymer charge vanishes and $k_D w\ll 1$  (thick EDL regime),
in the absence of pressure gradient the velocity at midpoint turns out to be
 \bea
 u_x(0)
 &\approx&
 -\frac{E_x\Sigma}{\eta \beta^2 w} \Bigl(1-\frac{\cosh(\beta h)}{\cosh(\beta w)}\Bigl)
 \nonumber\\
 &\approx&
 \begin{cases} 
 -\frac{E_x\Sigma}{\eta\beta^2 w}    & \mbox{if }   \beta h \gg 1 \\ 
  -\frac{E_x\Sigma}{\eta w}(w^2-h^2)  &  \mbox{if }   \beta w \ll 1 
 \end{cases}
 \label{vmidopint2}
\eea
which should be contrasted with the polymer-free case results $u_x(0)\approx-\frac{E_x\Sigma}{\eta k_D}$
and  $u_x(0)\approx-\frac{E_x\Sigma }{\eta } w$, in the case $\beta h \gg 1$ and 
$\beta h \ll 1$, respectively.  \end{itemize}
The majority of the existing studies \cite{cao2011controlling,cao2010electroosmotic,hickey2009molecular,tessier2006modulation} have focused on the small Debye length regime
  (molarities of the order 1M) ,
  corresponding to very thin EDL, because MD simulations at lower concentrations face severe problems
  of statistical accuracy due to the small number of ions considered.
  On the other hand, the LBM being based on the phase space distribution functions, does not run into
  such a difficulty and one can access the low concentration regime.
  We shall use the analytical method in the forthcoming section \ref{Numerics2} in order to interpret the numerical findings
  and provide a theoretical guide. 

\section{Numerical set-up}
\label{Numerics}
The evolution equations for the distribution functions are solved by employing the Lattice Boltzmann
method for ternary charged mixtures recently proposed by the authors
and based on the discretization of the kinetic equations on a discrete mesh.
 A detailed account   of the employed  numerical methods is provided in ref. \cite{monteferrante2014lattice} so that  we refrain from repeating the derivation in the present paper for space reasons.
The simulated system consists of a slab of fluid lattice points (nodes) enclosed by two  planes representing the channel walls. The lateral dimensions
of the channel, expressed in lattice units (l.u.), are $40\times 10$ and the channel width, $2 w$, is $250$ l.u..
The presence of grafted polymers is represented by two slabs adjacent the two walls each of thickness $\delta$ corresponding to
$50$ l.u.. Within these two regions $N_s$ fixed scatterers, with $N_s$ varying in the range $[0,100]$, occupy at random
the $40\times10\times 50$  cells.
Flow is in the $x$ direction and periodic boundary conditions are used in the $x$ and $y$ directions.
The walls are impenetrable to the fluid particles and no-slip boundary conditions are implemented by
using the bounce back prescription \cite{melchionna2011electro}.

Throughout this paper we shall use lattice units defined in the following.
Lengths are measured in units of lattice spacing
 $\Delta x=1$, the charge $e$ and the mass $m$  are assumed to be unitary, 
 the thermal energy $k_B T$ is specified by fixing the thermal velocity whose value is
$v_T=\sqrt{k_B T/m}=1/\sqrt{3}$ in l.u., the kinematic viscosity  of water $\nu$ is set equal to $1$ l.u.,
 while the dielectric constant is fixed
 by the  Bjerrum length $l_B=e^2/(4\pi\epsilon k_B T)$ and the length $\beta^{-1}$ varies from $[10,\infty]$ l.u. as $N_s$ varies between $100$ and $0$. 
To simulate the electro-osmotic flow it is necessary to resolve the EDL, so that $\lambda_D$ must be sufficiently larger than the lattice spacing.

 In physical units the lattice spacing is $\Delta x^{phys} =0.1 \, nm$,  $l_B=0.7 \, nm$.
 $\lambda^{phys}_D=2.3 \, nm$ (corresponding to a $18 \, mM$ aqueous solution),  
 $\nu^{phys}=10^{-6} m^2 s^{-1}$ and  the mass density of water $18 \, gr/cm^3$
  at room temperature .
 The physical magnitude of the  time step is obtained from  the relation:
\be
\nu^{phys} \frac{\Delta t^{phys} }{(\Delta x^{phys} )^2}=\nu
\ee
and by using a unit viscosity in lattice units, it follows that $\Delta t^{phys}  = 10^{-14} \,s$.

 As in ref. \cite{monteferrante2014electroosmotic}  we tune the parameters of our simulation 
 in order to match those of  
 the experiments where the inner walls of  a micro-channel  composed of a $SiO_2$ glass are coated  
 by Poly(DMA-GMA-MAPS)  and a buffer aqueous solution of ionic strength of 18 mM
 made of   $H_3PO_4-NaOH$ pH 2.5, 6- $\epsilon$-aminocaproic acid -
acetic acid pH 4.4, $H_3PO_4-NaOH$ pH 7.0, and Bicine-TRIS pH 8.5 fills the pore. 
  The concentration of the charge carriers is obtained by imposing the condition of global electro-neutrality for a given surface charge density and with the fixed value of the Debye length.

 In order to validate  our  theoretical estimate of the friction parameter $\gamma$ we 
 preliminarily simulated  a bulk periodic system of size $V_s=10\times40\times250$ l.u.
 having the same density, salt concentration
 and  obstacle density as those encountered in the channel case.
 Here the scatterers are distributed uniformly across the whole bulk simulation system.
   We applied a constant uniform force, $F$, everywhere, in the absence of surface charges and electric fields, and measured $\gamma$
with the help of the relation between the applied force and the resulting flow velocity according 
to the formula: $ \gamma= \frac{F}{m u} $. 
The measured $\gamma$ compares very well with the theoretical prediction:
\be
\gamma= k_p \frac{N_s}{V_s}
\ee
with $k_p=2$ l.u.,  where $k_p$ is the strength constant of the drag force  introduced in eq. \ref{dragforce}. 
By using $n_s=\frac{N_s}{V_s}=5 \times 10^{-3}$ l.u. we obtained a value $\gamma=10^{-2}$ l.u.
(corresponding to $\gamma^{phys} = 10^{10} \, s^{-1}$)
and $\beta=10^{-1}$ l.u.
(corresponding to $\beta^{phys} = 0.1$ $nm^{-1}$).

Concerning the polymer coating, we choose a polymer slab of  thickness $\delta=5$ nm, about twice the Debye length. 
In order to describe charged polymers as experimentally studied by several authors
\cite{chiari2000new,danger2007control,cao2012modulation}
we placed the idealized polymers in the regions adjacent to the wall. 
Following the choice of Monteferrante et al. \cite{monteferrante2014electroosmotic},
we varied $n_s$ from $0$ to $5\times 10^{-3}$ l.u..
   In order to assign such a  fractional  charge to each scattering center,  
   we desumed  the total number of polymers coating the walls and the charge of a single polymer from the experiments.
For the case of the polymer coated channel, the experimentally measured polymer densities ranged between $0.1$ and $0.18 \, gr/cm^3$. 
Thus to each of the
 $N_s$ sites randomly fixed in a volume $V_s=40\times 10\times 50$ near each wall, we assigned
 a charge fraction $z_s=10^{-4}$, so that the total charge of a polymer layer is
$Q_{p}=e z_s N_s$.

\section{Numerical results for the electro-osmotic currents}
\label{Numerics2}

\begin{figure}[htb]
\includegraphics[clip=true,width=8.0cm, keepaspectratio]{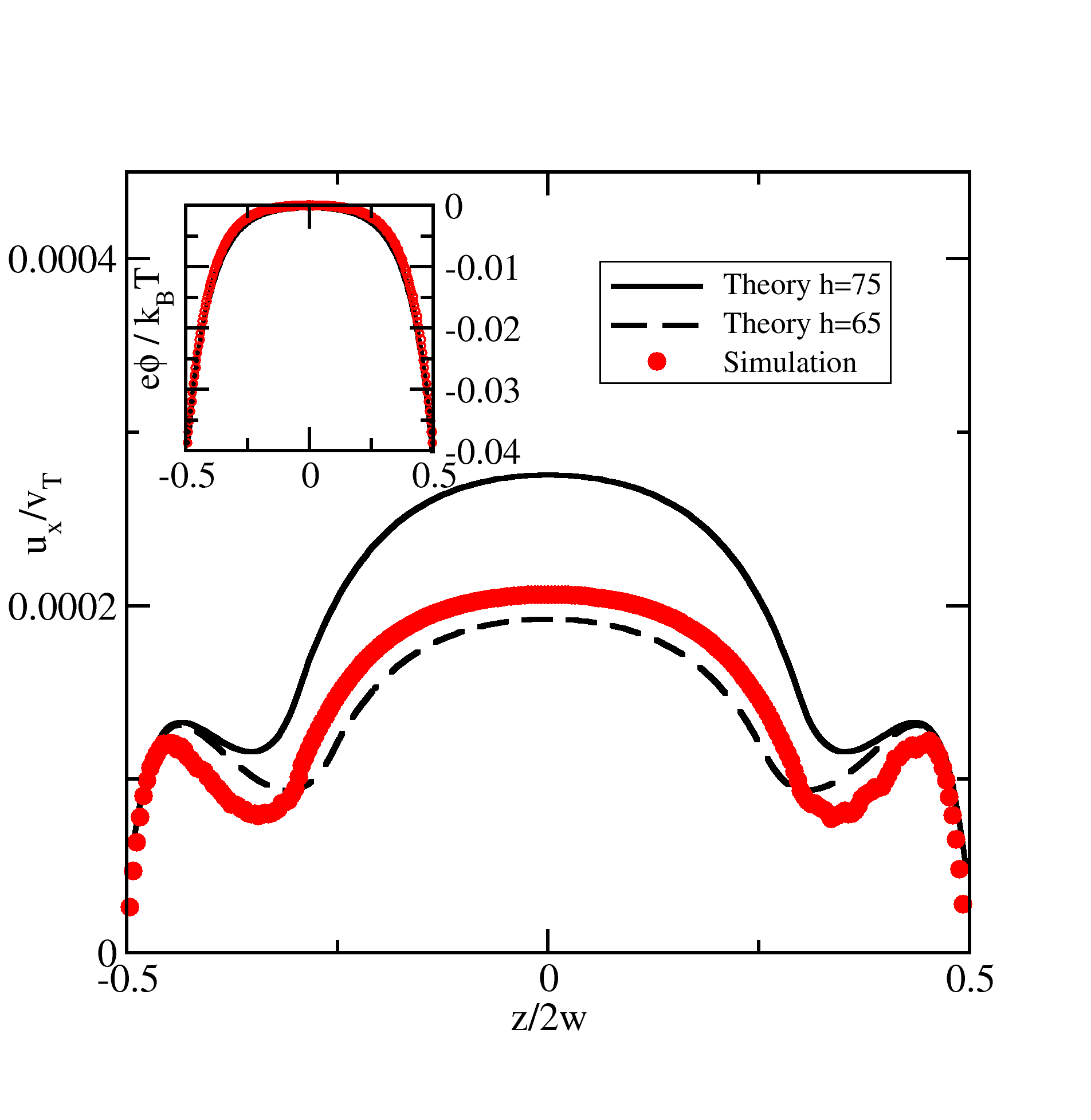}\\
\includegraphics[clip=true,width=8.0cm, keepaspectratio]{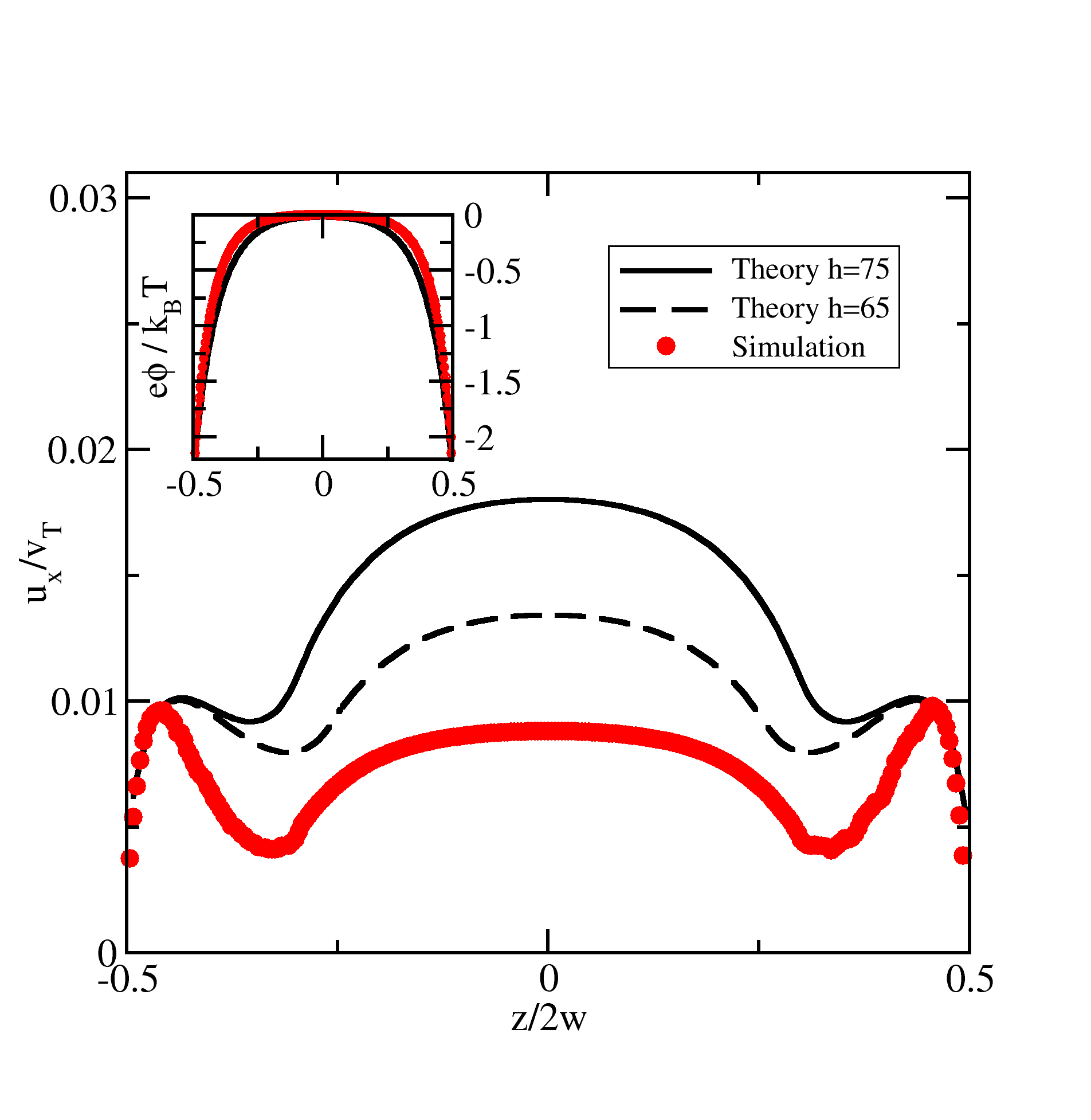}
\caption{Neutral polymer coating. Velocity profiles 
for $\beta=0.1 $ l.u. and two values of the surface charge density:
$\Sigma/e=-3.9\times10^{-5}$ (upper) and
$\Sigma/e=-1.3\times10^{-4}$ l.u. (lower panel).
The simulation curve (circles) is obtained for $2w=250$ and $h=75$ l.u.. 
The two theoretical curves are obtained for $2w=250$ l.u. and 
$h=75$ (solid) and $65$ l.u. (dashed), respectively.
Insets: electric potential profiles for each case. }
\label{figuno}
\end{figure}

We first consider 
the case of neutral polymers:
the LBM results  for the velocity and potential profiles are reported in fig. \ref{figuno} and 
compared to the predictions of the analytical theory of section \ref{Analytics}.
In the case of the  smaller surface charge density $\Sigma/e=-3.9\times10^{-5}$ l.u.,
the agreement is better than for $\Sigma/e=-1.3\times10^{-4}$ l.u. .  
In fact, the linearization
of the Poisson-Boltzmann eq. \ref{helmholtz} introduces a systematic error which becomes larger 
when $|\Sigma|$ increases,
as appreciated by comparing the two insets of fig. \ref{figuno}, where the analytical and numerical results
for the potential are compared. The discrepancy displayed in the right inset has repercussions on the
velocity profile, that is appreciably
lower in the central region in the LBM case. 
The reason for the discrepancy can be traced back to the small  gradient assumption used in the
constitutive equation for the stress tensor, eq. \ref{pressure}, which ultimately 
leads to eq. \ref{stokes1}.
Near the walls the velocity gradients are quite large and the Stokes equation might be 
inadequate to capture the correct behavior.
The difference between the analytic and the LBM  velocity profiles
decreases by setting $h=65$ l.u. in the analytical model and $h=75$ l.u. in the LBM.
However, such a readjustement is insufficient
to produce the same level of agreement in the case of larger surface charge, 
see fig. \ref{figuno}, right panel.
Probably one can improve the matching by adjusting both the value of $\Sigma$ and $h$
in the analytical model, but we did not pursue further
such a program because  somehow arbitrary, although it could provide a simple and economical  tool
to scan  the overall behavior of the system.

\begin{figure}[htb]
\includegraphics[clip=true,width=8.0cm, keepaspectratio]{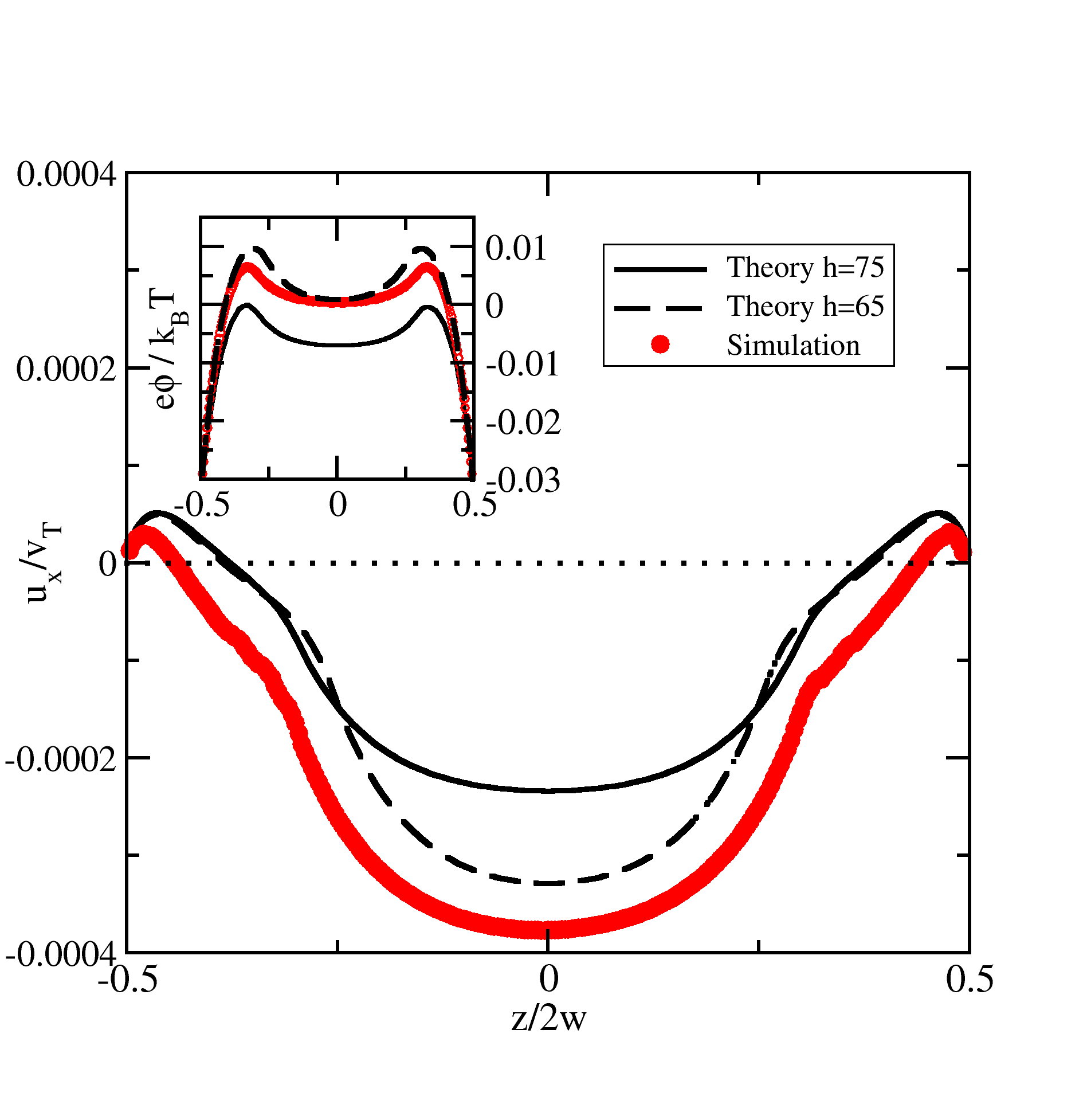} \\
\includegraphics[clip=true,width=8.0cm, keepaspectratio]{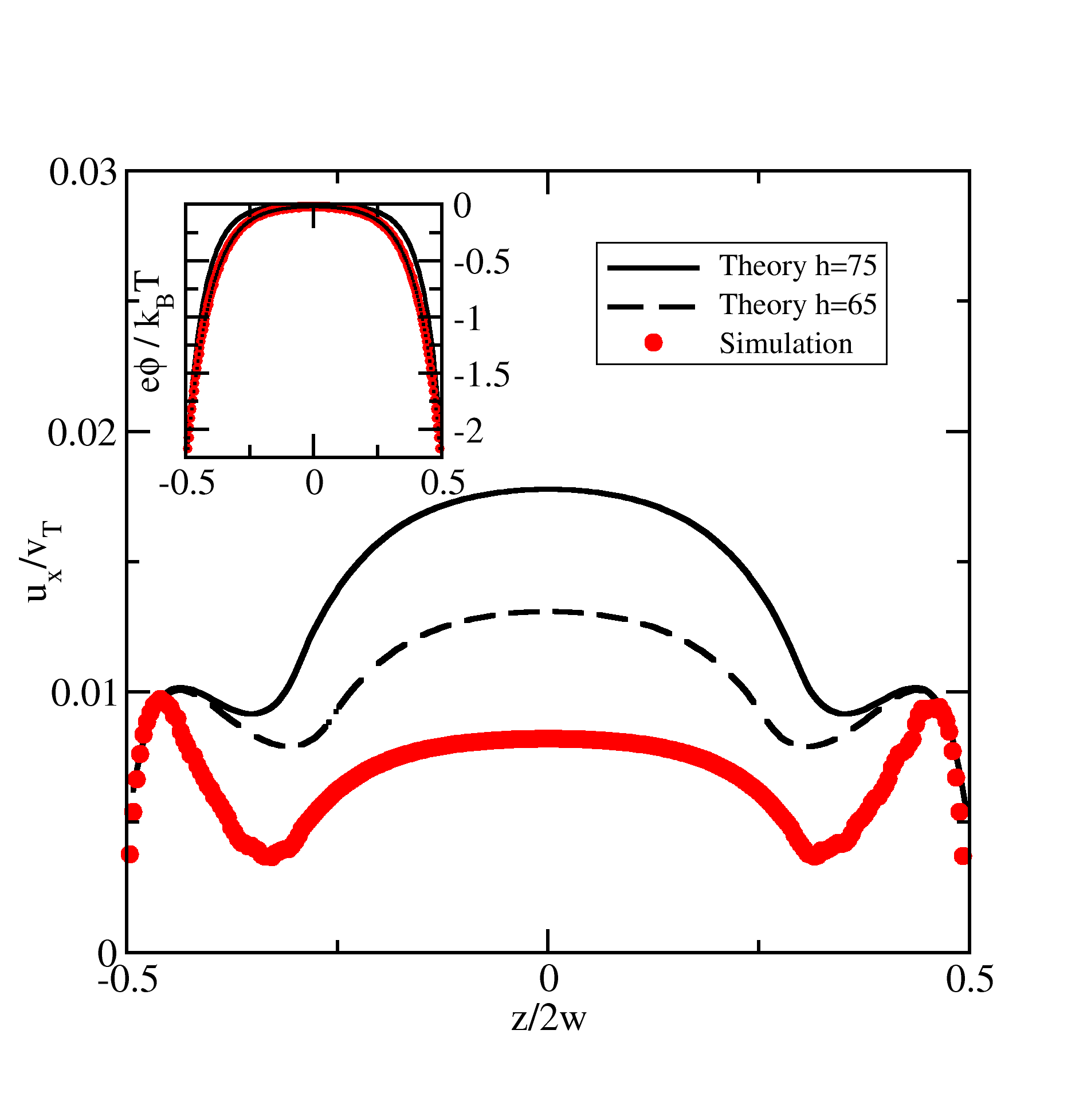}
\caption{Charged polymer coating. 
Velocity profiles for $\beta=0.1 $ l.u.   and $\Sigma/e=-3.9\times10^{-5}$ (upper)
and $\Sigma/e=-1.3\times10^{-4}$ l.u. (lower panel).
Symbols as in fig. \ref{figuno}.
 Insets: electric potential profiles. 
 Notice the flow reversal and the non-monotonic character of the potential  for
small surface charge density (left panel).
 }
\label{figdue}
\end{figure}

We consider, now, the case of charged coatings,
where
the electric potential depends altogether on the surface charge, the fixed charges associated with the fixed obstacles
and the ionic charges in the EDL. For low values of the negative surface charge it is possible to observe a flow reversal
in the presence of positive  polymer charges.
The density of mobile charges in the central region  $|z|<h$ has the same sign as the surface charge when $\Sigma$ 
is lower than the value
  \bea
 \Sigma^*  &=&
 \frac{e z_s n_s}{k_D }(
\sinh(k_D h)e^{-k_D w} - \sinh(k_D w) e^{-k_D h}) 
\nonumber\\
&\approx&
\begin{cases} 
 -\frac{e z_s n_s  }{2 k_D } e^{k_D \delta} & \mbox{if }   k_D\delta\gg 1 \\ 
- e z_s n_s \delta     &  \mbox{if }   k_D\delta\ll 1 \, ,
 \end{cases}
 \label{surfacefield}
  \eea
  as one can see from eq. \ref{explicitrep} and eq. \ref{plicitrep}.
      It is worth noting that the overall polymeric charge $z_s n_s \delta$ per unit surface does not need to be larger
      in absolute value  than the negative surface charge
 sitting on the wall in order to 
induce flow reversal in the capillary, consistently with the earlier observation 
by Hickey  et al. \cite{hickey2011influence}.
It is important to remark that the regime studied in most of the existing literature   \cite{tessier2006modulation,hickey2009molecular,
cao2010electroosmotic,cao2012modulation} concerns small values of the Debye
length, being typically $0.2\, nm$ . In this case, the EDL near the wall is not finely resolved and the comparison with the Smoluchowski theory is performed
by taking the peak position of the velocity as shear plane, where the velocity assumes its maximum value $v_{sl}$
and its derivative vanishes.
In our analytical model conducted at larger values of $\lambda_D$, instead, we can identify such a shear plane with the local maximum of the velocity,
 that appears in the region $h<|z|<w$, as a result of the competition between the drag force exerted by the polymers and the electro-osmotic
 force. Its location is at distance $\Delta\approx(k_D-\beta)\ln(\beta/k_D)$ from each wall and
 the velocity decays towards the bulk  from such a maximum in an exponential fashion. Only 
 for small values of $\lambda_D$ the velocity profile displays two side peaks and no dome at the center.
fig. \ref{figdue} displays the results relative to a weakly charged polymer coating, while the remaining parameters
 are identical to those employed in fig. \ref{figuno}. Again, for small surface charge the agreement between the 
 analytical result with the rescaled value of $h$ and the LB simulation is fairly good, but deteriorates at larger values of $\Sigma$
 for the same reasons discussed above.
 
 As displayed in  the inset of fig. \ref{figdue} left, the potential $\phi$ associated with a small negative surface charge, is non monotonic, reflecting
 the strong inhomogeneity  of the ionic charge distribution: the regions adjacent the walls are richer of counterions, whereas 
 further away the coions prevail. The direction of the mass flow occurs  in the $x$ direction in the first region
 and in the opposite direction in the negatively populated  region. Because of the relatively large value of the screening length 
 of fig. \ref{figdue} left one can see a large bulge in the center, while for  $\lambda_D  \ll  \delta$ the flow
 inversion takes place only at the walls . 
 
 We obtained   the solution of the Stokes eq. \ref{stokes1} 
 by imposing the continuity of the derivative of the velocity at $z=\pm h$.
 However, the latter condition is not necessary in principle,
 and has been introduced after observing that the LBM velocity profiles do not display cusps.
  Without such a continuity requirement, the analytic velocity profiles obtained by setting
 the parameter $N=0$ in  eq. \ref{vsolution} 
 display  lower values of the velocity in the central region, hence are closer to the
 LB result, but  also displays a cusp which is not observed numerically.

 The large $k_D w$ regime is of particular interest, because in the case of uncoated channels it displays
 plug-like velocity profiles: a very thin  region near the walls of thickness $\lambda_D$ where the velocity  raises from the
 zero value at the wall  to the plateau value $u_x(0)$. The effect of the obstacles is dramatic:
 one observes only a single peak near each wall and 
 the structure of the velocity profile depends weakly on  $k_D$, the relevant length being, now, $\beta^{-1}$. 
 If the polymer coating is sufficiently thick ($\beta\delta>1$),  besides the peaks near the walls
 the velocity drops exponentially towards the center, with the characteristic length $\beta^{-1}$.
 The peaks result from the competition between the electroosmotic driving  force, determining the growth
 of the velocity from the wall value (respecting the no-slip boundary conditions), and the antagonistic
 frictional force tending to suppress it.

\begin{figure}[htb]
\includegraphics[clip=true,width=8.0cm, keepaspectratio]{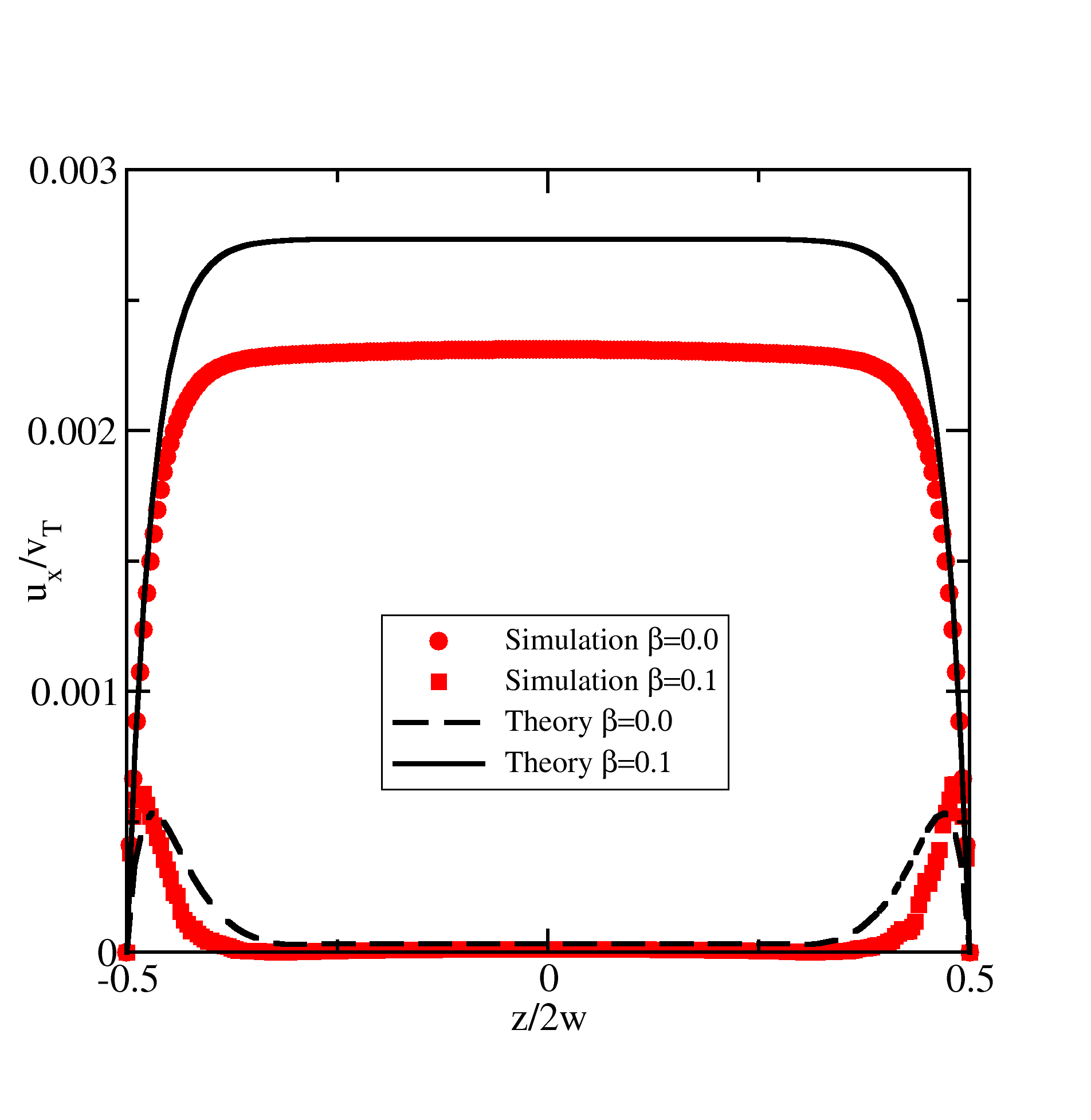}
\caption{Theoretical (black) and simulation (red) velocity profiles,  
for a slit of width $w=250$, polymer thickness $\delta=50$ and neutral polymer.
The curves correspond to $\beta=0$ (circles, solid) and $\beta=0.1$ l.u.
(squares, dashed), and
for $\Sigma/e=-1.3\times10^{-4}$ l.u. and $\lambda_D=7.5$ l.u.. }
\label{figprofilivsalfa}
\end{figure}

In this regime 
 the coating  appears
 to be very efficient in suppressing the mass flow, as shown in fig. \ref{figprofilivsalfa}.
  The continuity condition of the derivative of $u_x$ becomes irrelevant in the structure of the analytic solution since the cusp is hardly detectable and
  the difference between the solution with continuous derivative and discontinuous derivative 
  is very small.  We also remark that in the large $k_D$ regime, the 
  flow reversal is characterized by the presence of two 
   peaks near the walls, where most of the flow occurs, whereas in the small $k_D$ regime the majority 
  of the flow occurs at the center of the slit.
  
\begin{figure}[htb]
\includegraphics[clip=true,width=8.0cm, keepaspectratio]{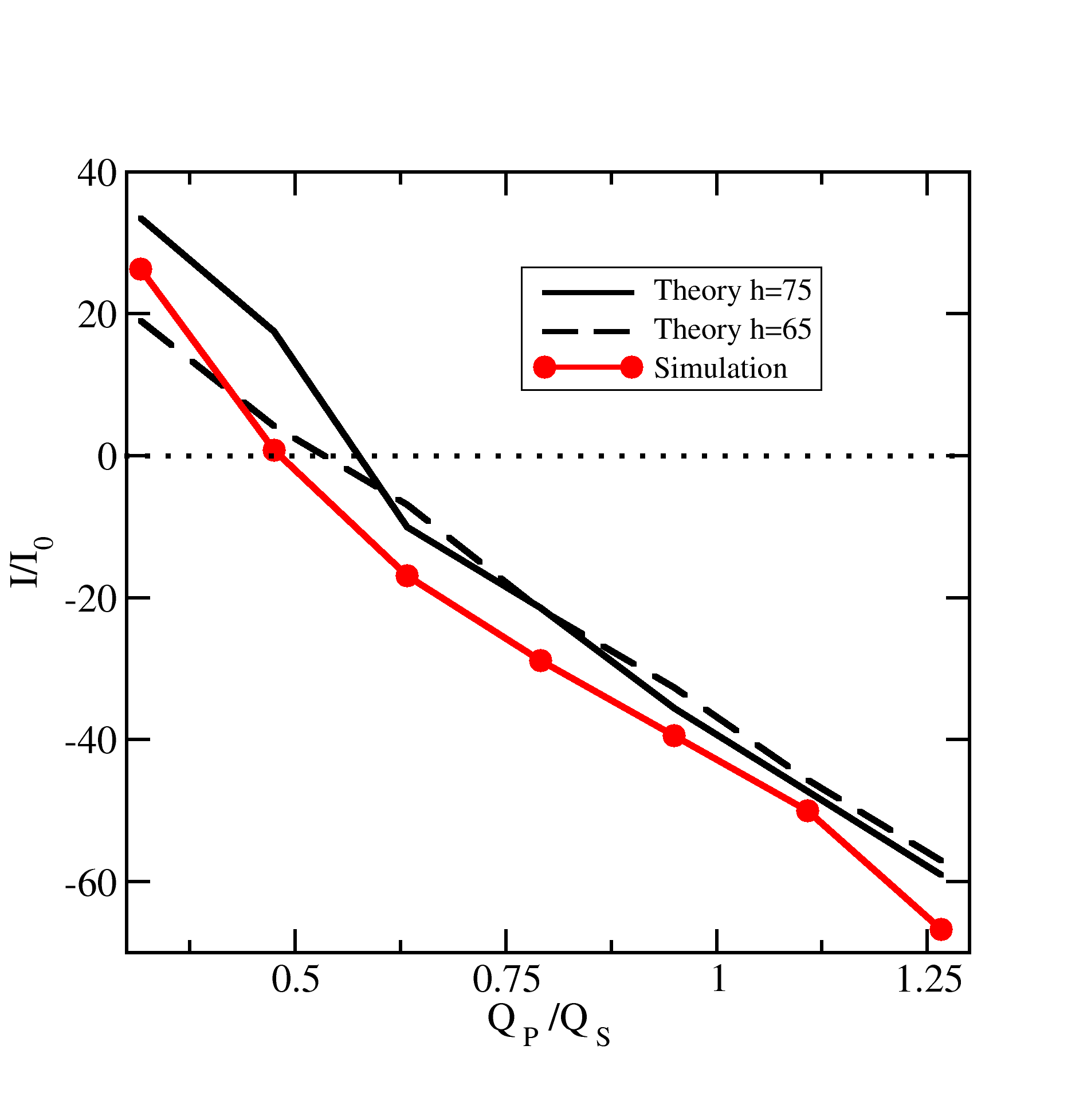} \\
\includegraphics[clip=true,width=8.0cm, keepaspectratio]{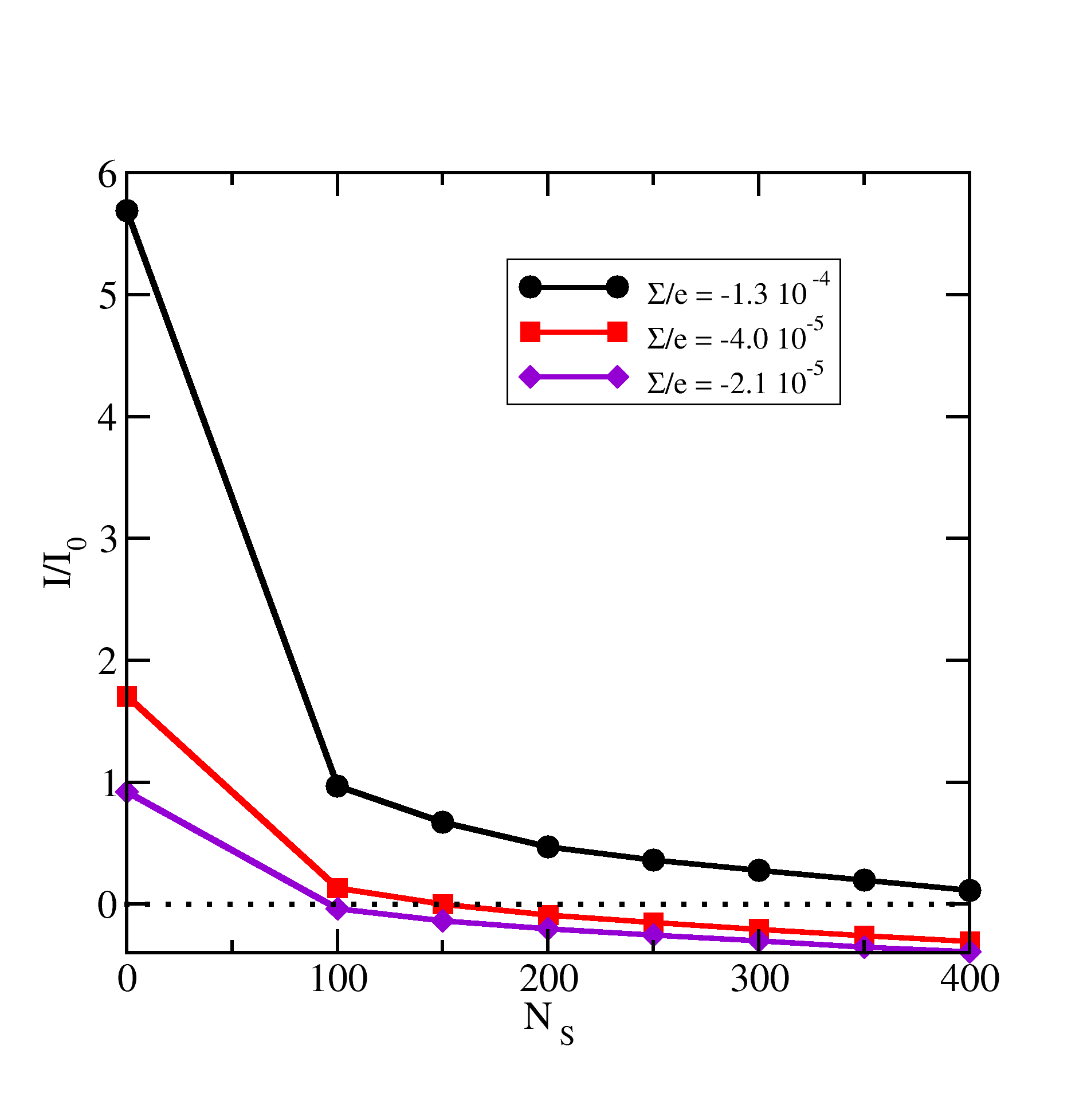}
\caption{Scaled mass flow rate ($I_0=E_x\Sigma L_x L_y/\nu$) 
vs the ratio between the total surface and polymer charge $Q_P/Q_S$ (upper) 
and vs the number of charged obstacles for different values of the surface charge 
(lower panel).
Data on the left panel correspond to simulation (red circles) and
theoretical results (black) for $h=75$ (continuos line) and $65$ l.u. (dashed line), respectively. 
Data on the right panel are from simulations and for $\lambda_D=23$ l.u..
Notice the flow reversal for the two lowest values of the surface charge.}
\label{figtre}
\end{figure}

 In fig. \ref{figtre} we report  the normalized mass flow rate as a function of the ratio between the total surface and total polymer charge.
 The theoretical and the simulation results are quite similar, and the difference between the 
 theoretical curves
 $h=75$ and $h=65$ l.u. is not large.  This  figure shows that the analytic prediction,
 as far as channel averaged quantity is concerned, is quite accurate 
 and can be used to give quantitative information about the global behavior.
 The right panel of fig. \ref{figtre} displays the mass flow, $\Phi$, versus the number of charged obstacles, $N_s$,
 for different values of the surface charge. Clearly, $\Phi$ decreases with increasing values of $N_s$,
 due to the increased friction.
 
\begin{figure}[htb]
\includegraphics[clip=true,width=8.0cm, keepaspectratio]{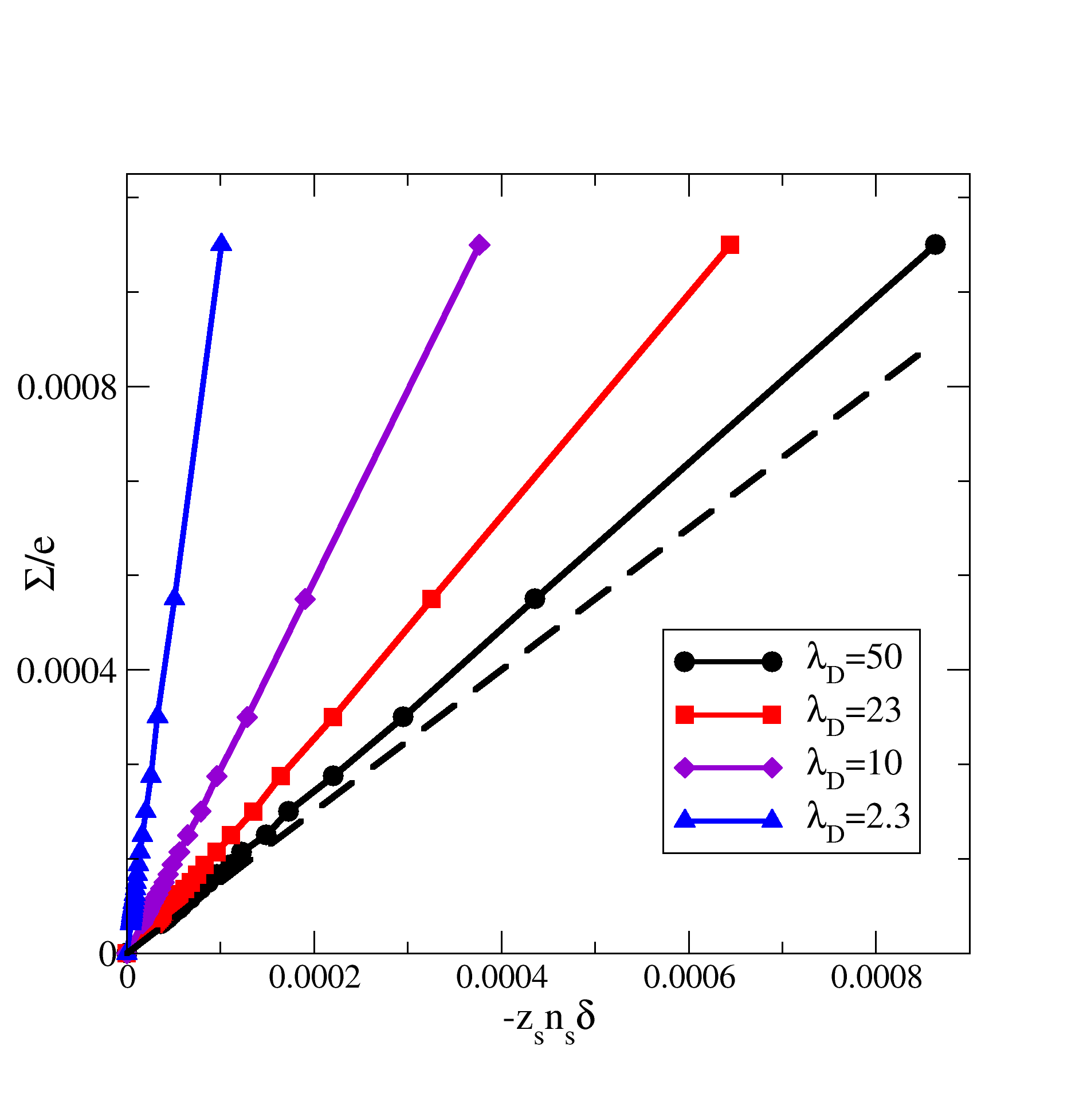}
\caption{Phase diagram. 
The four lines correspond to 
$\lambda_D=50$ (circles), $23$ (squares), $10$ (diamonds) and $2.3$ l.u. (triangles up).
The EOF is positive above each curve and negative below.
The dashed line represents the line $-\Sigma/e=z_s n_s \delta$
obtained when the charge in the polymer layers equals 
the surface charge (local electroneutrality assumption), 
a limit valid in the weak screening regime ($\lambda_D > \delta$).}
\label{figquattro}
\end{figure}

 In fig. \ref{figquattro}  we propose a phase diagram using the $\Sigma,(z_sn_s\delta)$ plane: 
 above the curve
 the flow is in the positive direction, whereas below the curve the flows is reversed.
 The different curves correspond to different Debye lengths expressed in lattice units:
 $\lambda_D = 2.3, 10, 23, 50$ (blue, green, red, black). 
The dashed line represents the curve $-\Sigma^* = z_sn_s \delta$ the electroneutral line, 
 that is, the line obtained by assuming that the charge in the polymer layer
 equals exactly the surface charge. Such a limit is attained in the weak screening regime ($\lambda_D  > \delta$).
 For all Debye lengths considered, the demarcation curve between direct and reverse flow shows
 a linear shape.
 Notice the dependence of the slope of the line on the Debye length: 
 the smaller $\lambda_D$ corresponds to 
 steeper lines since the surface charge is screened faster and produces a smaller effect.
 The larger $\lambda_D$ the bigger the polymer charge density where the flow reversal occurs, since the effects
of the surface extends over larger distances.

 Finally we comment on the charge current, while
 the mass current is approximately linear with respect to $\Sigma$,  the electric current is quadratic.
Whereas the mass current displays inversion for small values of the surface charge and positive adsorbed polymers, 
the charge current does not.  The reason is that the majority of the charge sits near the walls
where the velocity is also large. The reversal of  mass current is due to the fact that this is a flat average
of $u_x(z)$, while the charge current is an average of $u_x(z)$ weighted by the local charge of
carriers, and thus no inversion takes place.


\section{Conclusions}    
 \label{Conclusions}    
 In order to investigate the modulation of the electroosmotic flow in polymer coated capillaries
  we developed a
phenomenological model and described the grafted polymers by a set of  charged fixed obstacles
exerting both a Coulomb force and a drag force on the fluid species. 
The electrolytic solution is represented by a ternary mixture and its behavior is studied 
numerically.
By analyzing the resulting velocity profiles  we found that  the coating  
induces features that are not observed in standard EOF, such as non monotonicity,
velocity inversion, suppression of the plug-like profile in the small $\lambda_D$ regime. 
This opens the possibility of designing functionalized capillary surfaces in order to improve 
the resolution in capillary electrophoresis.
A remarkable feature of the model is that it lends itself to an analytical treatment 
in the case of moderate surface charges, so that the rather complex behavior in terms
 of a reduced set of  parameters can be predicted. 
 We have obtained the explicit analytical solution of the linearized model
 and shown that it agrees at semi-quantitative level with the numerical solutions.

 Our  approach replaces the complexity of the polymer layer by an assembly of  scatterers
 and thus neglects many important aspects such as the deformability of the polymers under the flow,
 their connectivity, or the possibility of forming mushroom or brush structures as discussed by Harden et al. \cite{harden2001influence} . Experimentally it would be important to establish a closer connection between our parameters $\delta$ and $\beta$ to the
degree of polymerization $N$ 
(related the  polymer thickness in the brush regime) and to  the number density of  grafted polymers onto  a flat surface, respectively,
instead of fixing them by a fitting procedure.
Another intriguing aspect that can be investigated by our methods  is the role played by
the charge distribution within the polymer layer on the resulting EOF.  
As an example, Danger et al. \cite{danger2007control}  used polyelectrolytes of different charge densities to control the EOF,
obtained by depositing a  first cationic polyelectrolyte layer  followed by depositing a second polyelectrolyte layer based on anionic
copolymer.

Before concluding, we remark that the differences between the planar and the cylindrical geometry 
are quantititatve rather than qualitative, so that it is possible
to extend the present approach to include such a geometry,  with an analytic treatment
slightly more involved and less transparent.


\section{Acknowledgments}
This work was supported by the Italian Ministry of University and Research 
through the ``Futuro in Ricerca'' project RBFR12OO1G - NEMATIC.
The authors wish to thank Marina Cretich, Marcella Chiari and Laura Sola
for insightful discussions.


\begin{figure}[htb]
\includegraphics[clip=true,width=8.0cm, keepaspectratio]{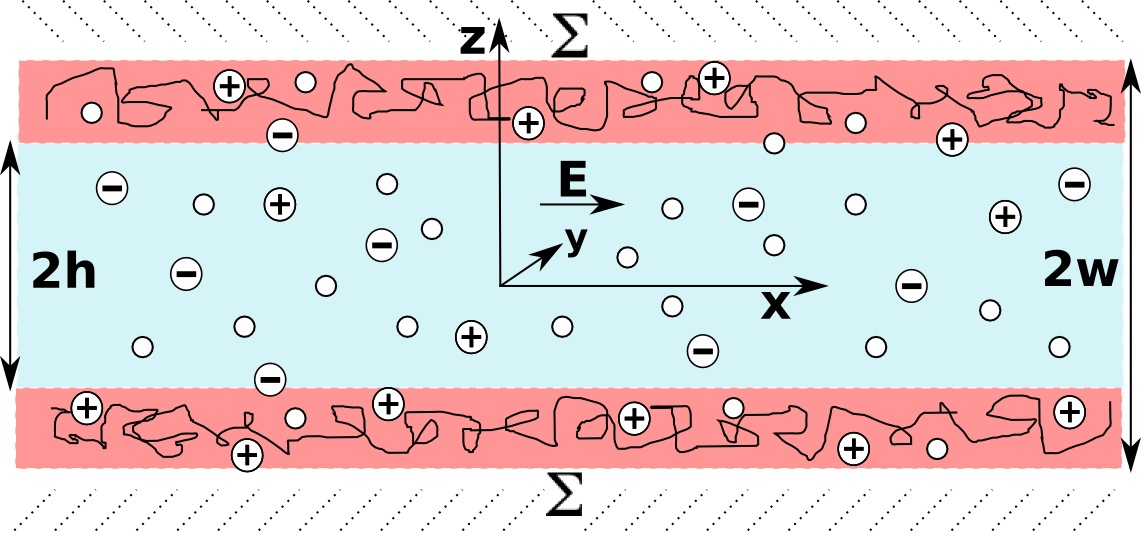}
\caption{Sketch of the pore system. The near-wall regions represent the polymer coating,
the small circles represent the solvent particles, the large circles represent counter and co-ions. }
\label{figurasistema}
\end{figure}


\bibliography{bib_moquette}
\bibliographystyle{rsc} 

\end{document}